\documentclass{acm_proc_article-sp}

\newdef{definition}{Definition}

\begin{document}

\title{Quantifying Reticulation in Phylogenetic Complexes Using Homology}

\numberofauthors{2}

\author{
\alignauthor
Kevin Emmett\\
       \affaddr{Department of Physics}\\
       \affaddr{Department of Systems Biology}\\
       \affaddr{Columbia University}\\
       \affaddr{New York, New York 10027}\\
       \email{kje2109@columbia.edu}
\alignauthor
Raul Rabadan\\
       \affaddr{Department of Biomedical Informatics}\\
       \affaddr{Department of Systems Biology}\\
       \affaddr{Columbia University}\\
       \affaddr{New York, New York 10032}\\
       \email{rr2579@cumc.columbia.edu}
}

\maketitle

\begin{abstract}
Reticulate evolutionary processes result in phylogenetic histories that cannot be modeled using a tree topology.
Here, we apply methods from topological data analysis to molecular sequence data with reticulations.
Using a simple example, we demonstrate the correspondence between nontrivial higher homology and reticulate evolution.
We discuss the sensitivity of the standard filtration and show cases where reticulate evolution can fail to be detected.
We introduce an extension of the standard framework and define the median complex as a construction to recover signal of the frequency and scale of reticulate evolution by inferring and imputing putative ancestral states.
Finally, we apply our methods to two datasets from phylogenetics.
Our work expands on earlier ideas of using topology to extract important evolutionary features from genomic data.
\end{abstract}




\section{Introduction}
\label{sec:introduction}
Evolutionary relationships are often depicted using trees.
From a topological perspective, trees have a simple structure, being contractible to a point.
However, several evolutionary processes involve the exchange of genetic material by mechanisms which cannot be modeled by tree.
These processes are collectively referred to as \emph{reticulate evolution}, examples of which include species hybridization, bacterial gene transfer, and homologous recombination.
As molecular sequence data accumulates, the importance of these processes has become increasingly apparent \cite{Gogarten:2005da}.
Here we expand on the use of ideas from topological data analysis, primarily persistent homology, to characterize reticulate evolution.

Persistent homology computes topological invariants from point cloud data \cite{Carlsson:2014cn}.
The application of persistent homology to molecular sequence data was introduced in \cite{Chan:2013vt}, where recombination rates in viral populations were estimated by computing $L_p$ norms on barcode diagrams.
In that paper, it was shown that persistent homology provides an intuitive quantification of reticulate evolution in molecular sequence data by measuring deviations from tree-like additivity.
While that approach has proved successful at capturing large scale patterns of reticulate evolution, the sensitivity for detecting specific reticulate events is lower.
This decreased sensitivity can be due to either incomplete sampling or weakly supported reticulations.
Here, we introduce an approach for imputing latent ancestors into the data that increases the quantitative signal detected by persistent homology.
Our approach is built on the \emph{median graph} construction.
Median graphs form the basis for a large number of phylogenetic network algorithms and are closely related to split decompositions of finite metrics \cite{Bandelt:1999wb,Bandelt:1992ko}.
A common desire is an approach to quantify the complexity of the resulting construction.
We show that using persistent homology of the median closure set is a fast and efficient way to characterize the phylogenetic incompatibility in the dataset.

The structure of the paper is as follows.
In Section~\ref{sec:background} we review the application of persistent homology to sequence data.
We present two simple examples in which the standard filtration fails to capture reticulation.
In Section~\ref{sec:median_complex} we introduce the median closure as an extended construction on the original vertex set.
We show how the persistent homology of this construction recovers quantitative signal of phylogenetic incompatibilitiy.
Finally, in Section~\ref{sec:examples} we present examples of our approach on two real sequence datasets.

\section{Persistent Homology for Sequence Data}
\label{sec:background}
In this section we briefly review the ideas in \cite{Chan:2013vt} as they relate to the application of persistent homology to sequence data.
Throughout, we assume biallelic data under an infinite sites model with no back mutation.

\subsection{Persistent Homology}
\label{subsec:persistent_homology}
Persistent homology computes topological invariants representing information about the connectivity and holes in a dataset.
A dataset, $S=(s_{1},\ldots,s_{N})$, is represented as a point cloud in an $l$-dimensional space, where $l$ is the length of the sequences.
From the point cloud, a nested family of simplicial complexes, or a filtration, is constructed, parameterized by a filtration value $\epsilon$, which controls the simplices present in the complex.
The standard filtration is Vietoris-Rips, in which a a simplex is present at scale $\epsilon$ if the pairwise distance between each element in $\sigma$ is less than $\epsilon$.
The filtration is represented as a list of simplices defined on the vertices of $S$, annotated with the $\epsilon$ at which the simplex appears.
Given a filtration, the persistence algorithm is used to compute homology groups.
The $0$-dimensional homology ($H_0$) represents a hierarchical clustering of the data.
Higher dimensional homology groups represent loops, holes, and higher dimensional voids in the data.
Each feature is annotated with an interval, representing the $\epsilon$ at which the feature appears and the $\epsilon$ at which the feature contracts in the filtration.
These filtration values are the \emph{birth} and \emph{death} times, respectively.
The topological invariants in the filtration are represented in a barcode diagram, a set of line segments ordered by filtration value on the horizontal axis.

\subsection{Evolution}
\label{subsec:vertical_evolution}
\begin{figure}
\centering
\includegraphics[width=\columnwidth]{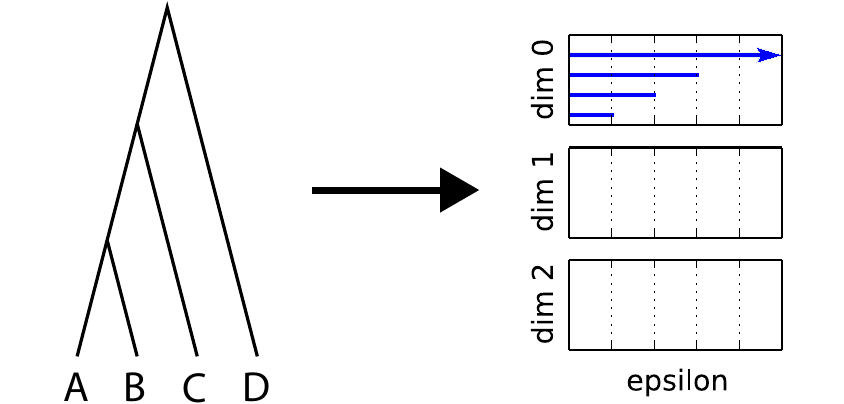}
\caption{A tree topology is contractible and will have vanishing higher homology, as reflected in the barcode diagram.}
\label{fig:tree_theorem}
\end{figure}

In the standard model of evolution, novel genotypes arise via mutation during reproduction.
In this case, evolutionary relationships will be accurately modeled as a bifurcating tree.
A tree is trivially contractible, and hence has vanishing higher homology (see Figure~\ref{fig:tree_theorem}).
This result was proven for sequence data in \cite{Chan:2013vt}.
What was not shown was the inverse statement, that vanishing higher homology implies tree-like evolution.

A simple test for the presence of reticulation is given by the \emph{four gamete test}.
The test states that the simultaneous presence of haplotype patterns 00, 01, 10, and 11 is incompatible with strict vertical evolution.
Failing the four gamete test provides direct evidence for reticulate evolution.
One way to quantify recombination in a set of sequences is the Hudson-Kaplan test, which counts the minimum number partitions required in the data such that within each partition all sites are compatible \cite{Hudson:1985wh}.
However, the Hudson-Kaplan test gives no further information about evolutionary relationships.

\begin{figure}
\includegraphics[width=\columnwidth]{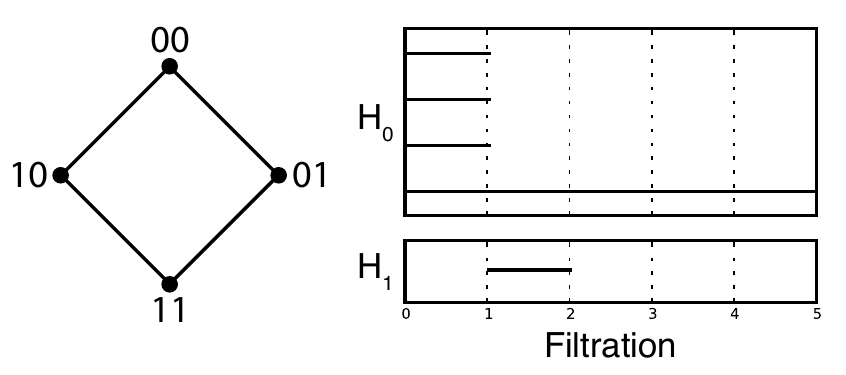}
\caption{The fundamental unit of reticulation. (A) The four gametes represent an evolutionary loop. (B) The barcode diagram with nonvanishing $H_1$ in the interval $[1,2)$.}
\label{fig:fundamental_loop}
\end{figure}

The four gametes can be considered the fundamental unit of recombination.
Topologically, this unit represents a loop, as shown in Figure~\ref{fig:fundamental_loop}.
Persistent homology identifies nonvanishing $H_1$ homology in the interval $[1,2)$.
We can give an interpretation to each vertex: there is a common ancestor, two parents, and a recombinant offspring.
In general, we do not \emph{a priori} know which sequences played which role in a given loop, effectively the same as the problem of rooting a phylogenetic tree.
Persistent homology is then simply a method for counting the number of such loops in the data, across all genetic scales.

In considering small examples of this form we often encountered situations in which the four gamete test indicated reticulate evolution, but persistent homology failed to detect a loop, as discussed in the two examples below.

\begin{figure}
\centering
\includegraphics[width=\columnwidth]{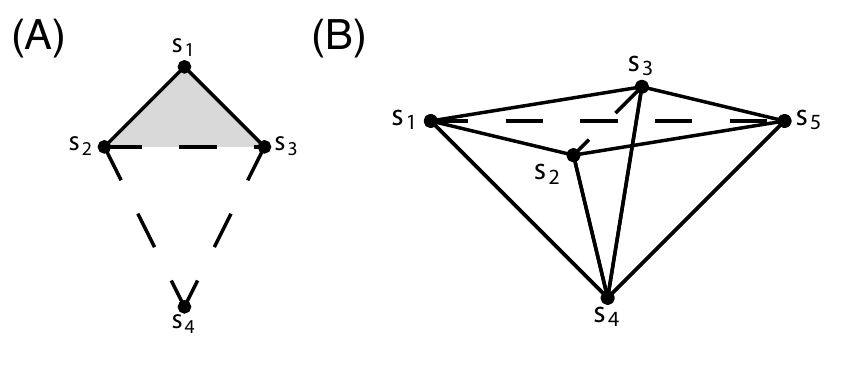}
\caption{Two examples in which the standard filtration fails to identify reticulate evolution. (A) In this example, the ancestral sequences collapse before forming a loop with the recombinant offspring. (B) In this example, multiple recombinations interact to create a degeneracy, and the entire complex collapses immediately. (From Song and Hein \cite{Song:2005vj})}
\label{fig:simple_examples}
\end{figure}

\emph{Example 1.} Consider the sequences $s_1=000$, $s_2=100$, $s_3=010$, and $s_4=111$.
The four-gamete test identifies incompatibility between sites 1 and 2.
However, persistent homology of the four sequences does not capture this reticulation.
To understand why, consider $s_1$ to be the common ancestor, $s_2$ and $s_3$ to be parents, and $s_4$ to be a descendant of a reticulate event.
In this scenario, we can infer that there was an ancestral recombinant sequence, $s_r=110$, which was not sampled.
The failure to find a loop is due to the ancestral and parent sequences collapsing before connecting with the recombinant offspring, as shown in Figure~\ref{fig:simple_examples}A.
In general, for a loop to be detected, the two internal distances must be greater than any of the four external distances.
In this case, the internal distance from parent 1 ($s_2$) to parent 2 ($s_3$), $d_{23}$ is equal to the distances from each parent to the sampled descendent of the recombinant ($d_{24}$ and $d_{34}$).
This is an example of incomplete sampling lowering the detection sensitivity, even in cases where incompatible sites are present.

\emph{Example 2.} This example is taken from \cite{Song:2005vj}.
Consider the sequences: $s_{1}=0000$, $s_{2}=1100$, $s_{3}=0011$, $s_{4}=1010$, and $s_{5}=1111$.
The four-gamete test identifies incompatibilities between sites $1$ and $3$, $1$ and $4$, $2$ and $3$, and $2$ and $4$.
Performing the Hudson-Kaplan test yields a partition between sites 2 and 3, however \cite{Song:2005vj} show a minimum of two reticulate events are required to explain the data.
Using the standard filtration, the complex contracts completely at $\epsilon=2$, and no higher homology will be detected.
In this case, the two reticulations interact in such a way that $s_3$ now sits equidistant from the other four sequences.
Had $s_{3}$ not been in the data, we would have had an example very similar to Example 1, with the interpretation of one recombination event.
In this example we observe that multiple reticulate events can interact in complicated ways, obscuring the signal from persistent homology.

\section{The Median Complex}
\label{sec:median_complex}
The median complex is an alternative construction on sequence data aimed at recovering signal of phylogenetic incompatibility using homology.
First, we define the median of a set of aligned sequences.

\emph{Definition 1.}
For any three aligned sequences $a$, $b$, and $c$, the \emph{median} sequence $m(a,b,c)$ is defined such that each position of the median is the majority consensus of the three sequences.

\begin{figure}
\centering
\includegraphics[width=.75\columnwidth]{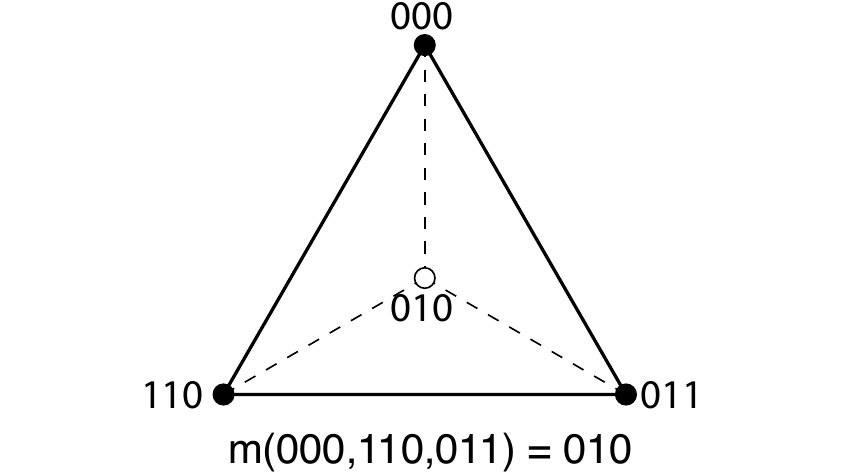}
\caption{The median is defined as the majority allele at each position. The median closure imputes the median into the original vertex set.}
\label{fig:median}
\end{figure}

Consider the example shown in Figure~\ref{fig:median}.
Here we have the three sequences $a=000$, $b=110$, and $c=011$.
Taking the majority allele at each position, the median is $m=010$.

Next, we define the \emph{median closure}.
Given an alignment $S$, the median closure, $\bar{S}$, is defined as the vertex set generated from the original set $S$ that is closed under the median operation,
\begin{equation}
\bar{S} = \{v \colon v=m(a,b,c) \in \bar{S} \:\forall\: a,b,c \in \bar{S}\}
\end{equation}
We obtain the median closure $\bar{S}$ by repeatedly applying the median operation to all sequence triplets until no new sequences are added.
The median closure consists of the original vertex set augmented by the computed medians.
We informally refer to topological complexes formed from the median closure as \emph{median complexes}.
We can then compute persistent homology on the new vertex set.

Filtrations on median graphs have been defined previously \cite{Dress:1997ch}, but not using explicit sequence representations.
To the best of our knowledge, quantification of the complexity of these objects has not been measured using homology.
We now revisit our two examples from Section~\ref{sec:background}.

\begin{figure}
\centering
\includegraphics[width=\columnwidth]{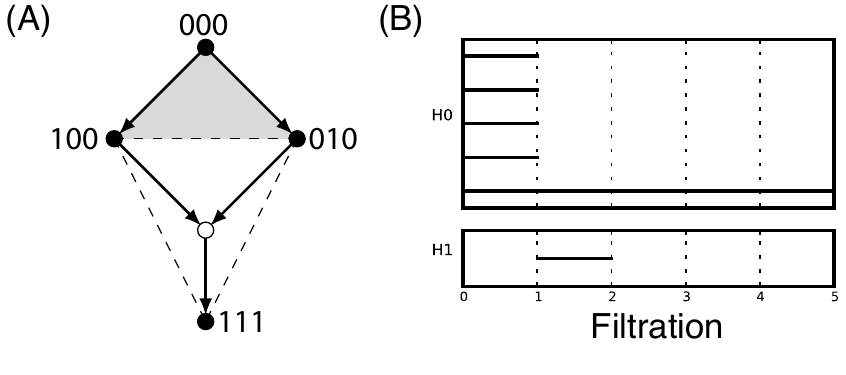}
\caption{One median node (white node), which acts as the recombinant offspring of $s_2$ and $s_3$. One $H_1$ loop detected in the interval $[1,2)$.}
\label{fig:example_1_revisited}
\end{figure}

\emph{Example 1.}
One median vertex, $m(s_2,s_3,s_4)=110$, as shown in Figure~\ref{fig:example_1_revisited}.
This vertex, labeled $s_r$, acts as the recombinant offspring of $s_2$ and $s_3$.
Persistent homology now detects an $H_{1}$ loop in the range $\epsilon=[1,2)$ formed between $s_1$, $s_2$, $s_3$, and $s_r$.
$s_4$ is interpreted the descendant of $s_r$.

\begin{figure}
\centering
\includegraphics[width=\columnwidth]{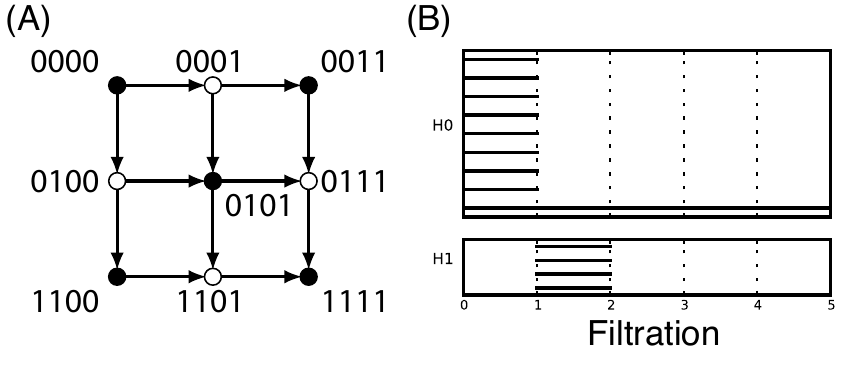}
\caption{Four median vertices (white nodes). Four $H_1$ loops now detected in the interval $[1,2)$.}
\label{fig:example_2_revisited}
\end{figure}

\emph{Example 2.}
Four median vertices, as shown in Figure~\ref{fig:example_2_revisited}.
Persistent homology now detects four $H_{1}$ intervals in the range $\epsilon=[1,2)$.
In this case, the median closure now overestimates the minimum number of recombinations required.
This example shows a potentially complicating aspect of the median closure in that specific $H_1$ features are no longer identifiable with specific reticulate events.

\section{Examples}
\label{sec:examples}
Here we consider two standard datasets from the phylogenetics literature.
In both examples, the standard filtration yielded no higher homology.
We generated the median closure and computed homology on that.
Datasets are represented using a triangle-free network construction, which approximates the computed homology.

\subsection{{\secit{D. melanogaster}} Data}
\label{subsec:kreitman}
\begin{figure}
\centering
\includegraphics[width=\columnwidth]{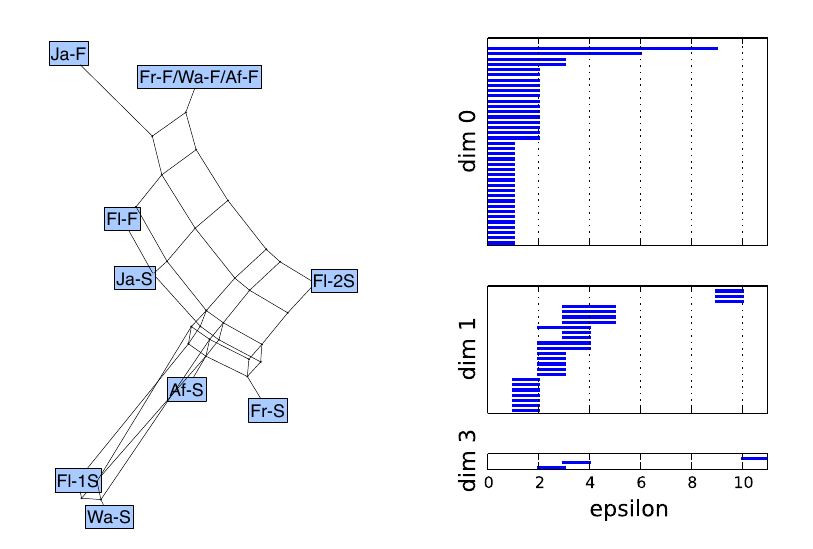}
\caption{Recombination in \emph{D. melanogaster}. Persistent homology identifies several complex reticulations in the population.}
\label{fig:kreitman}
\end{figure}

A benchmark dataset in studying recombination is the Kreitman data \cite{Kreitman:1983eq}.
The dataset consists of eleven sequences (nine unique) of the Adh locus from \emph{Drosophilia melanogaster} collected from various geographic locations, with 43 segregating sites.
The Hudson-Kreitman test yields 6 reticulate events.
Computing the median closure expands the dataset to 46 vertices.
Here we have non-trivial homology: 32 $H_1$ loops and 3 $H_3$ loops.
In the visualized network, the complex reticulations ($H_3$) are localized to the bottom-most samples.
The $H_1$ reticulations, on the other hand, are not very localized and persist across geographic regions.
The barcode plot is shown in Figure~\ref{fig:kreitman}.

\subsection{{\secit{Ranunculus}} Data}
\label{subsec:buttercup}
\begin{figure}
\centering
\includegraphics[width=\columnwidth]{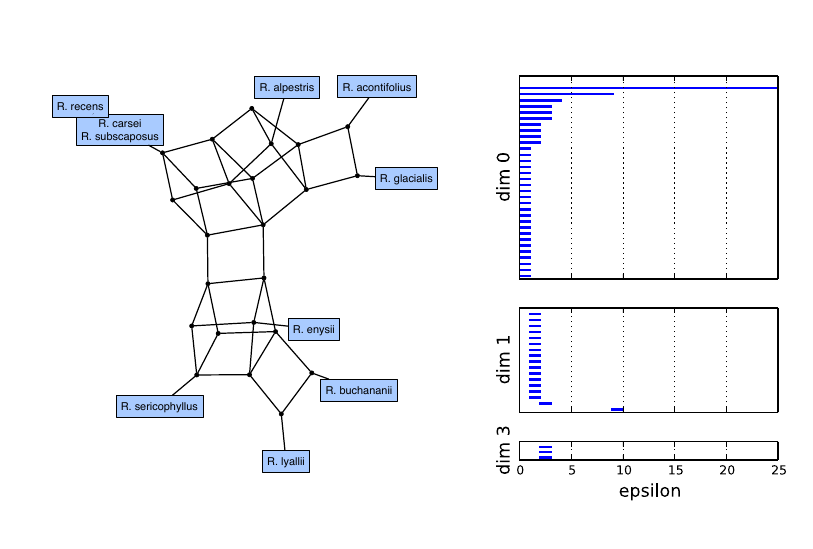}
\caption{Species hybrdiziation in genus \emph{Ranunculus}. Persistent homology identifies two populations separated by complex reticulations.}
\label{fig:buttercup}
\end{figure}

Natural hybridization occurs frequently in plants.
Here we examine reticulation in the maturase K (matK) protein in nine species from genus \emph{Ranunculus}.
This data is originally from \cite{Huber:2001vv}.
From nine initial species, the median closure has 32 vertices.
Persistent homology is computed and the barcode diagram shown in Figure~\ref{fig:buttercup}.
Looking at $H_0$, we identify two clusters of species.
Further, we identify 17 $H_1$ loops and 3 $H_3$ loops.
Comparing with the \emph{D. melangaster} data, reticulation at this locus is both smaller in scale (shorter bars at small filtration values) and less frequent (fewer total bars).
Additionally, the complex reticulations are localized within each $H_0$ cluster.

\section{Conclusions}
\label{sec:conclusion}
Persistent homology can capture and quantify complex patterns of reticulation in genomic data.
The standard Vietoris-Rips filtration is susceptible to reduced sensitivity due to incomplete sampling or interactions between reticulations.
Constructing the median closure of the original sequence set increases the topological signal of reticulation.
Future work will focus on efficient implementations of constructing this closure.

\section{Acknowledgments}
KE and RR are supported by NIH grants U54-CA193313-01 and R01-GM117591-01.
KE thanks Daniel Rosenbloom for useful discussions.

\bibliographystyle{abbrv}
\bibliography{topdrim4bio-MedianComplex}

\begin{thebibliography}{10}

\bibitem{Bandelt:1992ko}
H.-J. Bandelt and A.~W. Dress.
\newblock {A canonical decomposition theory for metrics on a finite set}.
\newblock {\em Advances in Mathematics}, 92(1):47--105, 1992.

\bibitem{Bandelt:1999wb}
H.-J. Bandelt, P.~Forster, and A.~R{\"o}hl.
\newblock {Median-joining networks for inferring intraspecific phylogenies.}
\newblock {\em Molecular Biology and Evolution}, 16(1):37--48, 1999.

\bibitem{Carlsson:2014cn}
G.~Carlsson.
\newblock {Topological pattern recognition for point cloud data}.
\newblock {\em Acta Numerica}, 23:289--368, May 2014.

\bibitem{Chan:2013vt}
J.~Chan, G.~Carlsson, and R.~Rabadan.
\newblock {Topology of Viral Evolution}.
\newblock {\em Proceedings of the National Academy of Sciences},
  110(46):18566--18571, Nov. 2013.

\bibitem{Dress:1997ch}
A.~Dress, K.~Huber, and V.~Moulton.
\newblock {Some variations on a theme by Buneman}.
\newblock {\em Annals of Combinatorics}, 1(1):339--352, Dec. 1997.

\bibitem{Gogarten:2005da}
J.~P. Gogarten and J.~P. Townsend.
\newblock {Horizontal gene transfer, genome innovation and evolution}.
\newblock {\em Nature}, 3(9):679--687, Aug. 2005.

\bibitem{Huber:2001vv}
K.~T. Huber, V.~Moulton, P.~Lockhart, and A.~Dress.
\newblock {Pruned median networks: a technique for reducing the complexity of
  median networks}.
\newblock {\em Molecular Phylogenetics and Evolution}, 19(2):302--310, 2001.

\bibitem{Hudson:1985wh}
R.~R. Hudson and N.~L. Kaplan.
\newblock {Statistical properties of the number of recombination events in the
  history of a sample of DNA sequences}.
\newblock {\em Genetics}, 111(1):147--164, 1985.

\bibitem{Kreitman:1983eq}
M.~Kreitman.
\newblock {Nucleotide polymorphism at the alcohol dehydrogenase locus of
  Drosophila melanogaster}.
\newblock {\em Nature}, 304(5925):412--417, Aug. 1983.

\bibitem{Song:2005vj}
Y.~S. Song and J.~Hein.
\newblock {Constructing minimal ancestral recombination graphs}.
\newblock {\em Journal of Computational Biology}, 12(2):147--169, 2005.

\end{thebibliography}

\balancecolumns

\end{document}